\documentclass[%
 reprint,
superscriptaddress,
 amsmath,amssymb,
 aps,
floatfix,
]{revtex4-2}
\usepackage{graphicx}% Include figure files
\usepackage{dcolumn}% Align table columns on decimal point
\usepackage{bm}% bold math
\usepackage{xcolor}
\usepackage{soul}
\usepackage{float}
\usepackage{amsmath}

\begin{document}

\preprint{APS/123-QED}

\title{Virtual-Diagnostic-Based Time Stamping for Ultrafast Electron Diffraction}

\author{F. Cropp}
\email{ericcropp@physics.ucla.edu}
\affiliation{Department of Physics and Astronomy, UCLA, Los Angeles, California 90095, USA}
\affiliation{Lawrence Berkeley National Laboratory, Berkeley, California 94720, USA}

\author{L. Moos}
\affiliation{Special Circumstances, 113 Cherry St. 94153, Seattle, Washington 98104, USA}

\author{A. Scheinker}
\affiliation{Los Alamos National Laboratory, Los Alamos, New Mexico 87544, USA}

\author{A. Gilardi}
\affiliation{Lawrence Berkeley National Laboratory, Berkeley, California 94720, USA}

\author{D. Wang}
\affiliation{Lawrence Berkeley National Laboratory, Berkeley, California 94720, USA}

\author{S. Paiagua}
\affiliation{Lawrence Berkeley National Laboratory, Berkeley, California 94720, USA}

\author{C. Serrano}
\affiliation{Lawrence Berkeley National Laboratory, Berkeley, California 94720, USA}

\author{P. Musumeci}
\affiliation{Department of Physics and Astronomy, UCLA, Los Angeles, California 90095, USA}

\author{D. Filippetto}
\email{dfilippetto@lbl.gov}
\affiliation{Lawrence Berkeley National Laboratory, Berkeley, California 94720, USA}

\date{\today}% It is always \today, today,

\begin{abstract}
In this work, non-destructive virtual diagnostics are applied to retrieve the electron beam time of arrival and energy in a relativistic ultrafast electron diffraction (UED) beamline using independently-measured machine parameters.  This technique has the potential to improve temporal resolution of pump and probe UED scans. Fluctuations in time of arrival have multiple components, including a shot-to-shot jitter and a long-term drift which can be separately addressed by closed loop feedback systems. A linear-regression-based model is used to fit the beam energy and time of arrival and is shown to be able to predict accurately behavior for both on long and short time scales. More advanced time-series analysis based on machine learning techniques can be applied to improve this prediction further. 
\end{abstract}

\maketitle

%\tableofcontents

\section{Introduction}

% Virtual diagnostics
A recent trend in accelerator and beam physics has been the use of virtual diagnostics to measure indirectly one or more beam parameters using larger sets of upstream, non-destructive measurements of accelerator and machine parameters, which are correlated with the downstream beam properties. Various mathematical tools ranging from linear and non-linear interpolations to more complex machine-learning based techniques can be used to create high fidelity predictive models from training data obtained by destructive beam measurements. Then the model can be used to retrieve the beam parameter of interest once the destructive measurements cease. This is critically advantageous when measurements of the given parameter are particularly time-consuming or require running in a particular working point on the beamline which is not compatible with the end-user application (e.g. \cite{SCHEINKER2020163902, emma2018machine,convery2021uncertainty,yakimenko2019facet,hanuka2021accurate,scheinker2015adaptive}).

% Smart control systems and global feedback 
In particular, in systems where beam fluctuations strongly affect the accelerator performances, a very attractive opportunity exists to take advantage of virtual diagnostics models to improve the reliability in delivering a known set of beam parameters to an application even in presence of active feedback systems. This is because while feedback systems can be used to monitor machine parameters and keep them close to a given working point, these loops are not perfect (i.e. still allow a residual amount of jitter) and a complete compensation requires a beam-based diagnostic. In addition, the monitoring is typically limited to a single variable and the control algorithm does not take into account cross-correlation terms with other machine parameters. A global smart control system taking advantage of powerful and reliable virtual diagnostics models has therefore the potential to outperform such local feedback loops.  

% UED case
As an example, temporal stability is particularly important in pump-probe ultrafast techniques such as Ultrafast Electron Diffraction (UED).  In a UED experiment, temporal resolution is defined as: 
\begin{equation}
    \tau=\sqrt{\Delta t_{e^-}^2+\Delta t_{laser}^2+\Delta t_{jitter}^2+\Delta t_{VM}^2},
    \label{eq:UED}
\end{equation}
where $\Delta t_{e^-}$ is the electron bunch length, and $\Delta t_{laser}$ is the laser pulse length.  These quantities can be reduced using a bunching cavity and laser compressor, respectively. $\Delta t_{VM}$ is the velocity mismatch term, which can be neglected for ultra-relativistic beams and thin samples.  That leaves the limiting factor of $\Delta t_{jitter}$, the time-of-arrival jitter between the laser pulse and electron bunch. Time-stamping has been proposed in the past (e.g. \cite{scoby2010electro, THzStamping}) to sort the UED patterns and retrieve the actual temporal trace of an ultrafast process. Nevertheless, depending on the particular implementation, accurate time-stamping strongly constrains the machine setup (charge, crystal proximity, THz deflector) which might not be fully compatible with high-quality diffraction patterns. Taking advantage of a virtual diagnostic would greatly increase the range of applicability of time-stamping in UED, potentially improving the temporal resolution of the technique. 

% Beam energy stability
One of the beam parameters most strongly connected to the time-of-arrival of the beam at the sample is the beam energy. In linear transport theory, the connection is mathematically represented by the matrix element $R_{56}$ which connects the final time-of-arrival with the relative energy deviation $\Delta E/E$ from the reference particle.  For example, in a drift, higher energy particles arrive sooner.  With more complex arrangements of beamline elements, which include buncher cavity and bending dipoles, the relation can become more complex (see \cite{filippetto2016design}), as discussed in Section~\ref{sec:energystamping}. Still, beam energy is often the dominant contribution to the particle time-of-arrival at a given plane in the beamline and a kinetic energy virtual diagnostics could be useful to refine the predictions of the relative time-of-arrival fluctuations in a UED setup \cite{zhao2018terahertz}. We also note here that there are other cases where non-destructive measurements of the beam energy (which otherwise require bending the beam in a dipole spectrometer) would greatly improve accelerator performances. For example, in multi-shot measurements of transverse phase spaces, such as a quadrupole or solenoid scan emittance measurements \cite{minty2003measurement,prat2014four} energy fluctuations change the focusing strength of the magnets, which would be considered to be constant for such a scan; poor energy stability is catastrophic to such a measurement. Even single-shot emittance measurement techniques, such as \cite{marx2018single} require knowing the beam energy.  

% In this paper (frame paper) 
In this paper, we develop non-destructive virtual diagnostics for the beam time of arrival (TOA) and kinetic energy which take into account non-destructive machine parameters measured upstream. 
%% new Daniele Sept 21
Time stamping techniques in UED pose unique challenges, requiring single-shot TOA measurements on very low charge beams with very high resolution (\textless 100 fs), which has only been achieved via destructive measurements \cite{filippetto2022ultrafast}.  Our work shows that the use of advanced mathematical methods can help breaking the paradigm of measurement accuracy versus beam charge. Indeed, we show that the electron beam parameters can be inferred from the accelerator context, i.e. measurable instantaneous machine parameters, with the same level of precision obtained by performing destructive measurements. Thus, the temporal resolution becomes independent from the beam charge, and only dependent on the precision of the measurement of machine parameters.

% In this paper...
The experiments were carried out at the LBNL HiRES beamline for UED, where we were able to reconstruct the beam energy and TOA for each shot with an accuracy beating our feedback systems 
using simple linear interpolation virtual diagnostics model. By applying machine learning (ML) techniques, the reliability of the prediction further improves. The application of ML has been shown to solve or mitigate a plethora of accelerator control and diagnostic problems, for example, for navigating efficiently the multi-dimensional parameter space to find control set points \cite{edelen2020machine, duris2020bayesian}, for inverting a large parameter space to make a parasitic diagnostic \cite{li2018electron,kabra2020mapping}, or for non-destructive virtual diagnostics \cite{SCHEINKER2020163902, emma2018machine}.  ML has also been combined with model-independent adaptive feedback for automatic control of the longitudinal phase space of the electron beam in the LCLS \cite{scheinker2018demonstration}.  Further, UED has benefited from ML-based static models and virtual diagnostics \cite{zhang2021accurate,zhang2022toward}. 

% Outline paper
In the next section, we discuss the operations at HiRES and the measurement systems for the parameters that are used in establishing the virtual diagnostics. The results of a linear-regression-based virtual diagnostic are shown in two cases where beam TOA and beam energy are used to train the model and benchmark its fidelity. In the last section of the paper, we compare the application of more complex ML models to the linear regression model.

\section{Synchronous Data Acquisition and Analysis at HiRES}

\subsection{Data Acquisition at HiRES}

\begin{figure*}[ht]
    \begin{center}
  \includegraphics[scale=1.25]{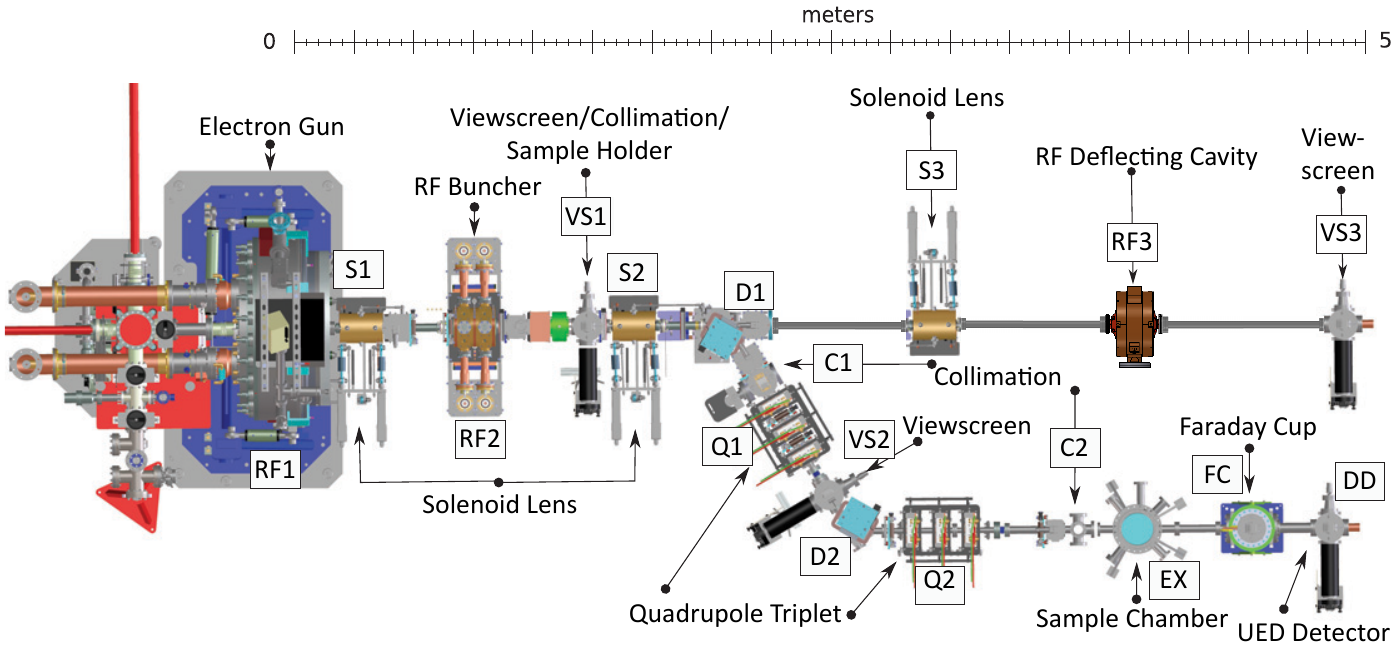}
  \end{center}
  \caption{The HiRES beamline.  The UED beamline starts at D1 and goes through the dogleg to DD, while the diagnostic beamline goes straight from D1 to VS3. Adapted from \cite{filippetto2016design}.}
  \label{fig:HiRES}
\end{figure*}

% Description of HiRES beamlines
The HiRES accelerator includes a continuous-wave-class, normal-conducting electron photogun working at 185.7 MHz \cite{sannibale2012advanced} (RF1 in Fig.~\ref{fig:HiRES}) and a subsequent bunching cavity (RF2) operating at the 7th harmonic of the gun, i.e. 1.3 GHz. The present maximum electron beam repetition rate is fixed by the photocathode laser to 250 kHz, while an acousto-optic deflector at the end of the optical amplification chain can select user-defined patterns and/or lower the repetition rate. The nominal beam energy is 750 keV and all measurements in the paper were taken with an approximate beam charge of 15 fC.

Referring to Fig. \ref{fig:HiRES}, a dipole magnet (D1) downstream of the gun (RF1) and and RF buncher cavity (RF2) is used to select between two beamlines, each providing access to a series of diagnostic tools. In particular, a deflecting cavity along the straight line (RF3) provides accurate pulse length and time of arrival information, with calibration of 23.37~fs/pixel at the downstream imaging screen (VS3). The side beamline (UED beamline), branches off at an angle of 60$^\circ$ with respect to the straight line, resulting in high dispersion at the imaging screen VS2, and enabling high resolution energy measurements. The energy calibration $\Delta E/E$ at the screen is $2.5\times 10^{-5}$/pixel, and can be increased or decreased using the quadrupole triplet just upstream VS2 (Q1).  In the measurements presented in this paper, the calibration for $\Delta E/E$ was $1.7\times 10^{-5}$/pixel. 

% Benefits of HiRES (UED & R&D)
Owing to its unique set of beam parameters and its flexibility, the HiRES has been used for both UED applications~\cite{durham2020relativistic,siddiqui2021ultrafast}, and for developing new technologies for compact and large-scale user facilities~\cite{ji2019ultrafast, scheinker_adaptive_2021, sannibale_high-brightness_2019}.

% Describe PID loops and FPGA, short term stability
For example, the low-level-RF control electronics (LLRF), one of the most critical sub-systems for ensuring electron beam energy stability, has been developed at LBNL and then deployed at the LCLS-II accelerator at SLAC~\cite{huang_low_2016}. The system allows precision control and measurement of amplitude and phase, with minimal cross-talk (more the 100 dB isolation in the upgraded version) and white noise background below 150 dBc/Hz. Such development is a key component for developing high precision feedback loop controls, and for high fidelity prediction of beam parameters.

% Synchrony figure
\begin{figure}[ht] 
    \begin{center}
  \includegraphics[scale=0.35]{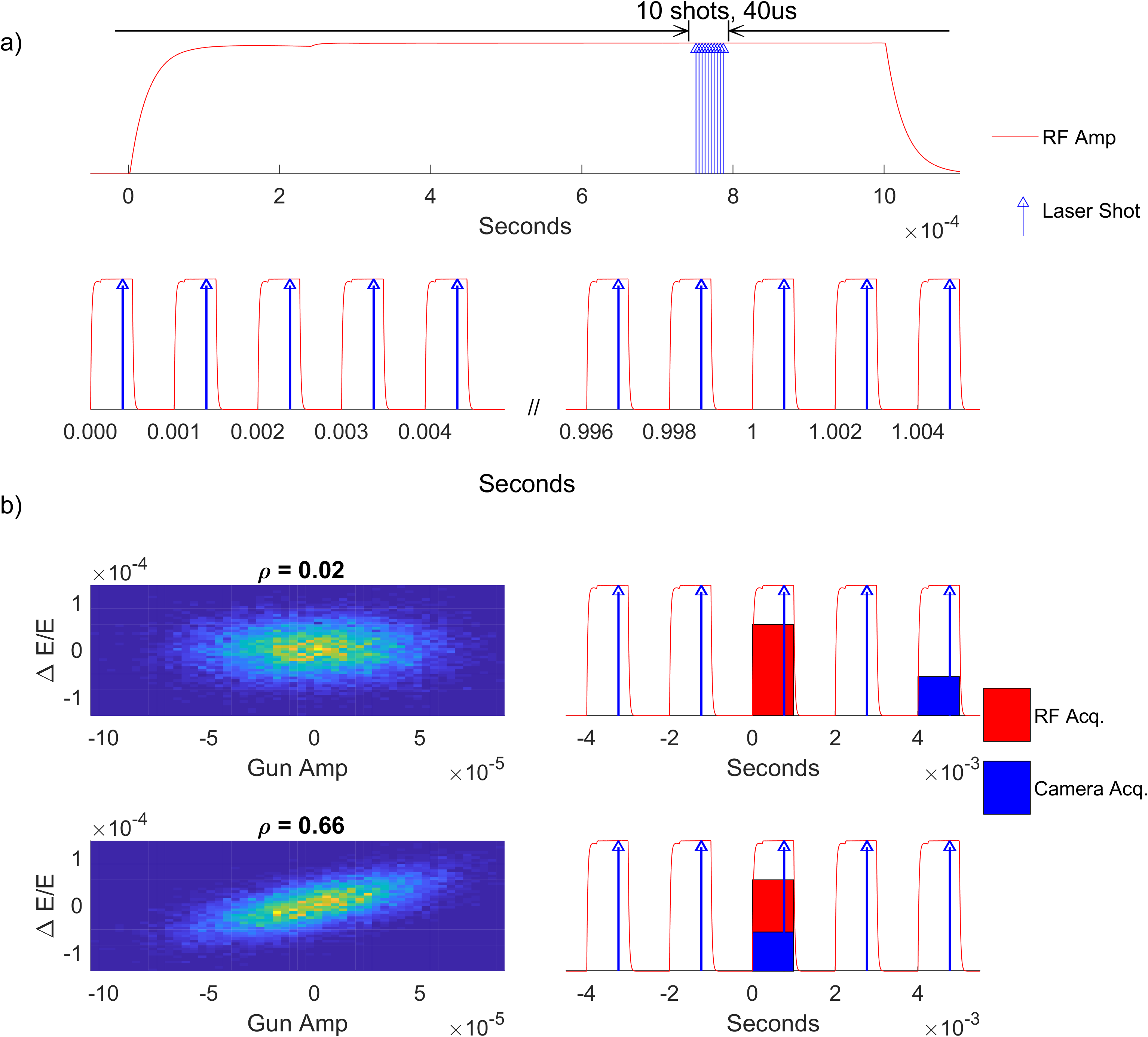}
  \end{center}
  \caption{a) Synchronization scheme: RF amplifier run at 50\% duty cycle with a 2~ms period.  Up to 10 laser shots arrive at 4~$\mu$s intervals for a short period toward the end of each RF pulse, where the RF is generally most stable.  b) Correlation plots of RF gun amplitude and relative energy deviation (measured at the dipole spectrometer) for synchronized and temporally misaligned acquisition schemes.}
  \label{fig:Synchrony}
\end{figure}

% Describe general data-acquisition scheme
The aim of this work is to develop a novel virtual diagnostic tool for online high-precision time and energy stamping. The tool has been tested for single-shot beam predictions using information collected passively during beam runs and, while in these first test the acquisition was limited to 1~Hz, minor modifications to the timing system would allow much faster repetition rates, in the kHz range and beyond. 

The development of a virtual diagnostic starts with building a model of the system, correlating measurements of beam parameters with machine parameters.  Measuring the beam energy or time of arrival requires intercepting the electron beam with scintillator screens and analyzing the resulting images for the beam centroid after bending through a dipole or a time-dependent kick from a transverse deflecting cavity (TCAV).

The SNR of the training datasets is of particular importance as the model will be trained on the processed variables extracted from the images, and any error in the calculation for the parameter corresponds to an effective loss of information. In order to boost image SNR, it is possible to integrate multiple electron beam pulses (because of the high repetition rate of the system, each only 4~$\mu$s apart), so long as the timescale of system changes is longer than the averaging period. 

At fixed RF power, the rate of change of the phase or amplitude of an electromagnetic field in a resonant cavity is limited by the cavity bandwidth. Indeed, the latter acts as a filter for external disturbances, so that every noise component outside its bandwidth is strongly attenuated. In the case of our 186 MHz CW-RF gun, with a quality factor $Q$ greater than $10^4$, we can estimate the time scale of field fluctuations:

\begin{equation}
   \tau_{noise}=\frac{1}{\Delta f}=\frac{Q}{f}>50 \mu s
    \label{eq:reciprocal_bandwidth}
\end{equation}
where $f$ is the resonance frequency of the cavity. Further, there is an intra-pulse PID-type feedback system engaged, which should further reduce jitter.  The engagement of the intra-pulse feedback can be seen after approximately 250~$\mu$s in the RF traces in Fig. \ref{fig:Synchrony}.  

For the RF bunching cavity and deflecting cavity, RMS fluctuations of the amplitude and phase of the RF in both cavities shows only a minimal increase when integrated for 40~$\mu$s relative to the case when integrated for 4~$\mu$s (the inherent uncertainty in laser shot time of arrival).  The slight increase in fluctuations in each cavity is expected to increase the uncertainty in beam time of arrival at the final screen by less than 10 fs each.  Therefore, most of the data produced in this work has been collected by averaging 10 beams per image, in order to increase the signal-to-noise ratio (SNR) in the images (see Fig.~\ref{fig:Synchrony}). 

% Ensuring synchrony
In order to obtain the most accurate model and predictions, the heterogeneous data acquired (a mix of images and waveforms) requires deterministic time-alignment with a precision equal or better than $ \tau_{noise}$. Figure~\ref{fig:Synchrony} describes our timing setup. The electron gun is used in pulsed mode for these experiments, with a total duration of the RF pulse of 1 ms and a repetition rate of 500 Hz (corresponding to a duty cycle of 50\%). The optical gate sending the burst of 10 consecutive laser pulses can be activated at any time along the RF pulse, with a precision of 4$\mu$s. In Fig. \ref{fig:Synchrony}b we show the correlation plots of electron beam relative energy deviation and the amplitude of the field in the electron gun, in the simple case where no other cavity is used. The $\rho$ coefficient on top of each plot corresponds to the value of the correlation function between the two. The data is shown in the case of synchronized and not synchronized acquisition, providing a clear idea of the information lost without precise time alignment. 

% Asynchronous justification

The beam position on the screen measured during the initial characterization of the accelerator is not only determined by the variable we are interested in predicting. Beam position can change because of a magnet current change, or because of laser pointing fluctuations on the cathode. Therefore a complete model should calculate correlations with all the relevant parameters of the accelerator. Photocathode laser beam images are saved with $\mu$s-scale alignment precision synchronously with electron beam images, but fortunately not all data requires the same level of time-alignment. Variations in parameters such as cavity temperatures, water flows and magnet currents mostly contribute to machine drifts, and only require synchrony at the sub-second level, which can be achieved via software. At HiRES, continuous data storing of user-requested machine setting is performed automatically by an online database with 10 Hz periodicity,  providing the necessary information to include all machine parameters in the model. 

\begin{figure}[h!]
    \begin{center}
  \includegraphics[scale=0.4]{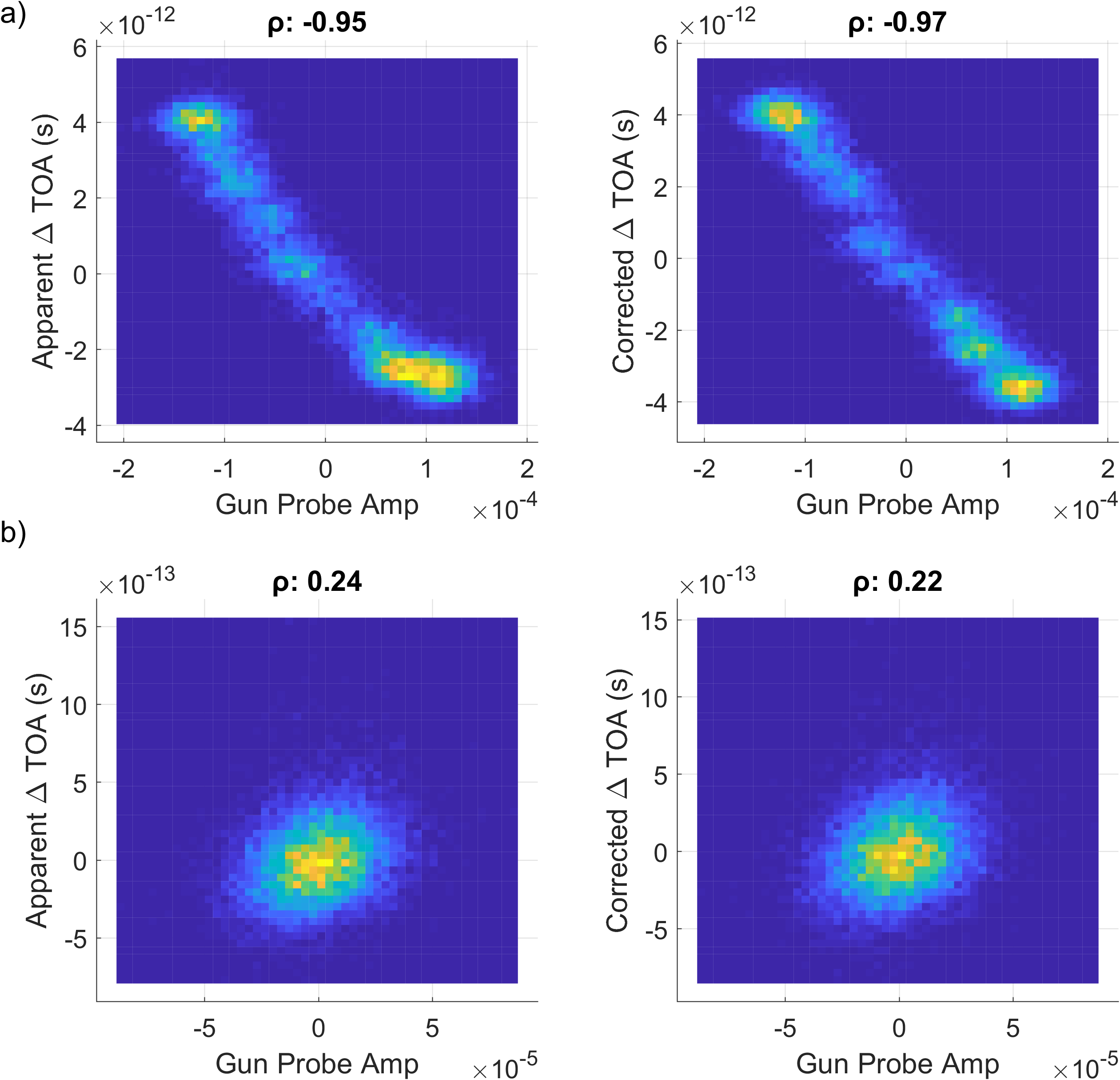}
  \end{center}
  \caption{Synchronous measurements of beam TOA at TCAV: correlations of TOA with electron gun amplitude (without the use of the RF bunching cavity, similar to data in Fig.~\ref{fig:Synchrony}).  Left: TCAV RF jitters are not taken into account in post processing.  Right: fluctuations in TCAV RF amplitude and phase are used to correct beam time-of-arrival measurements. In a) long term drifts are uncompensated.  In b) a moving average is subtracted to show only short time scale jitters. Note the much smaller y-scale in the bottom plots.}
  \label{fig:Async_Justify}
\end{figure}

Data for the deflecting cavity deserves a special discussion. As mentioned above, data on this cavity should in principle be taken synchronously, as small short-term fluctuations are expected. In Fig.~\ref{fig:Async_Justify} the correlation plots between the electron gun amplitude and the measured beam time-of-arrival before and after the compensation of TCAV short-term fluctuations are shown.  
Given the small contributions of these jitters to the measured short-term fluctuations, we acquire TCAV data via the database, and therefore only account only for long-term drifts of the field in the cavity.

% Asynchronous justification plot

\subsection{Data Analysis and Prediction}
\label{sec:linearmodel}
% Introduce linear regression
In the following section, virtual diagnostics based on multi-variable linear regression are presented.  Multiple linear regression is a statistical modeling technique where 
\begin{equation}
    \label{eq:LR1}
    \textbf{Y}=\textbf{X} \boldsymbol{\beta}+\boldsymbol{\epsilon}
\end{equation}
where $\textbf{Y}$ is a vector the observed quantities, $\textbf{X}$ is a matrix of dimension number of observations by number of predictors, $\boldsymbol{\beta}$ is a vector of regression coefficients (dimension: number of predictors) and $\boldsymbol{\epsilon}$ is the error term, a vector of individual errors on each observation.  Estimates of regression coefficients, $\boldsymbol{\hat{\beta}}$ are learned by minimizing residuals.

% Setup Sensitivity Analysis 
Multiple linear regression is a powerful tool because of its explainable nature.  Rather than producing a black-box model, which produces predictions through an opaque process, multiple linear regression produces an interpretable and explainable model. As it will be shown below, the regression coefficients are learned and can be analyzed in order to quantitatively characterize the impact that each predictor has on the overall prediction. 

% Discuss time independent methods
It is also important to observe that linear regression between two variables is agnostic to the time relation between different data points as it is inherently a time-independent method.  However, the data sets in this work are all time-series datasets. While linear regression is effective in quantifying the consistent effect that the predictors have on the observation, the method will fail to identify the time-dependent noise processes that perturb the system and affect both the predictors and the observations.

% Time dependent procedure
In this case, two adaptations to the usual prescription for linear regression were made: 1) the data were not randomized, in order to preserve the time-series ordering and 2) the model is trained on the first part of the data, while the last part of the data is reserved for validation.  The linear regression results below serve as a practical baseline result that -- by itself -- shows promise for improving stability, but as shown in the last section of the paper could be further improved upon by taking into account temporal evolution using more complex models.

\section{Online Predictions of Electron Beam Parameters}
\label{sec:onlinePred}
\subsection{Time Stamping}
\label{sec:timestamping}
% Characterize stability

The temporal resolution in UED experiments (see Eq. \ref{eq:UED}) is often dominated by relative time of arrival fluctuations between excitation laser and probing electron beam. Therefore an online diagnostic capable of precise non-destructive measurement of TOA would have a profound impact on the overall instrument performance.
For the data presented in this section, the accelerator setup matched the beam and machine parameters used during UED experiments. As such, the RF bunching cavity (RF2) is set for temporal compression, with nominal field amplitude and zero degree injection phase (the so called \textit{zero-crossing} phase). The fields in both the electron gun and the bunching cavity are stabilized in amplitude and in phase by fast, FPGA-based PID-type feedback loops.

\begin{figure}[h!]
    \begin{center}
  \includegraphics[scale=0.55]{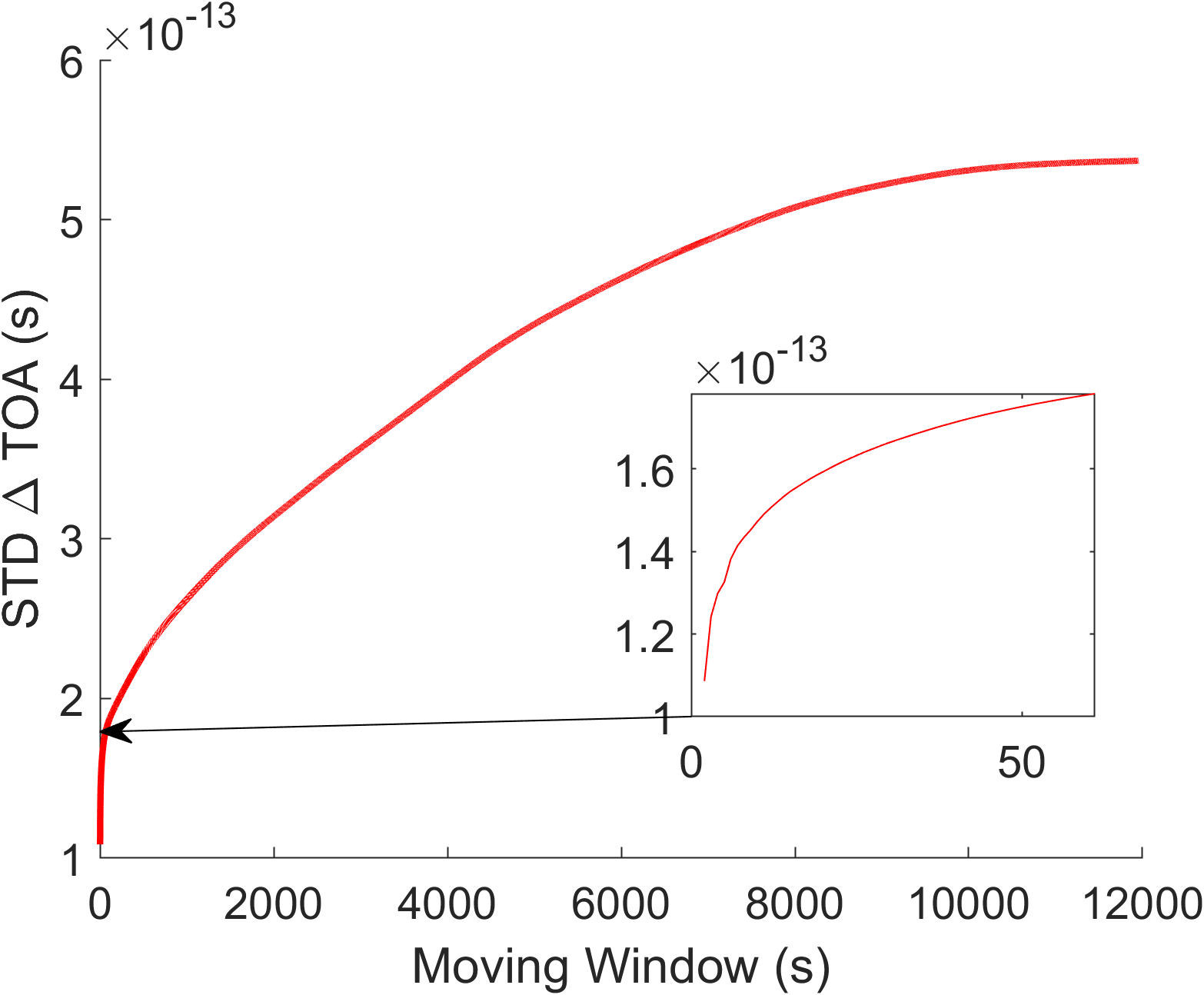}
  \end{center}
  \caption{Standard deviation of transverse beam centroid calibrated to TOA relative to the reference beam as a function of the width of the acquisition time window. Inset shows a short timescale.}
  \label{fig:Stability}
\end{figure}

In Fig. \ref{fig:Stability} we show the standard deviation of the time-of-arrival of the beam at the deflecting cavity (measured converting the centroid variation of the beam on the screen using the pixel to time deflector calibration), as a function of the temporal duration for the data acquisition. This quantity continuously increases due to short and long term drift. In the inset, zooming in on the 1 minute time scale, it is shown how short-term drifts account for less than 200 fs of temporal jitter. 
On the other hand, increasing the temporal width of the acquisition window the overall stability of the system is observe to degrade at longer timescales. Depending on the duration of the intervals in between re-establishing a new time-zero position in UED pump probe scans~\cite{filippetto2022ultrafast}, the integrated resolution can become as large as 600 fs.

% Plot to characterize stability

% Case 2 Results
\begin{figure}[h!]
    \begin{center}
  \includegraphics[scale=0.5]{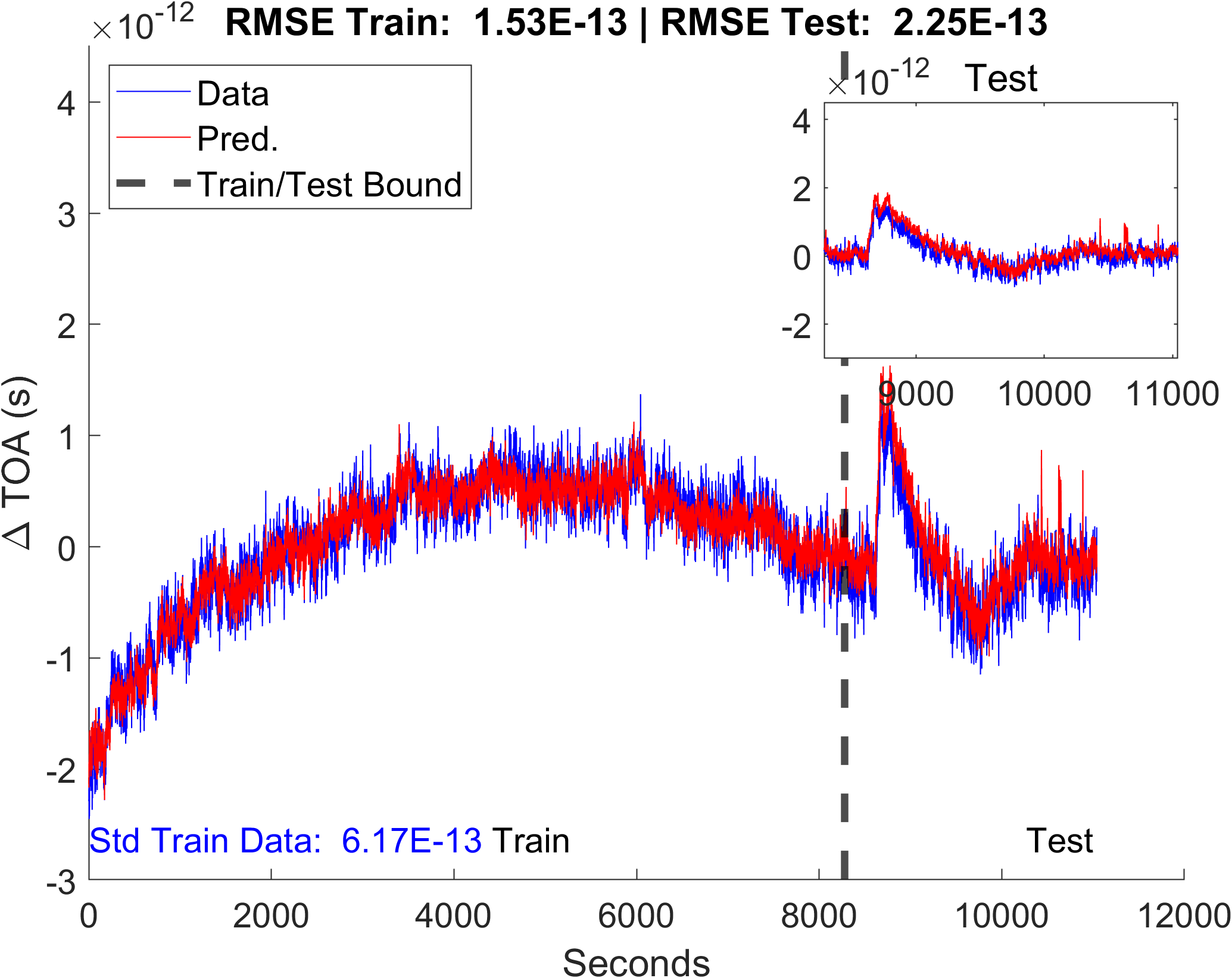}
  \end{center}
  \caption{Time of arrival fluctuation measured using the TCAV screen while the PID-type feedback was engaged. Residual drifts not correted by the feedback are present. The linear regression is also shown. The uncertainty due to long-term drifts is reduced to 200 fs level similar to the shot-to-shot, short-term jitter shown in Fig. \ref{fig:Stability}.}
  \label{fig:Case2}
\end{figure}

Figure~\ref{fig:Case2} shows the evolution of the beam TOA at the TCAV over about 3 hours. The data are divided in two sections (highlighted by the vertical dashed line), with 75\% of the points used for developing a model of the system (the training data set), and the last 25\% is used to validate it (test set).

% Discuss predictors
We then use the linear regression model described in Sec.~\ref{sec:linearmodel} for prediction. Inputs to the model include amplitude and phase of the electron gun, the bunching cavity and the deflecting cavity, photocathode laser arrival time at the cathode with respect to the RF wave and an image of its transverse shape, intensity and position at the cathode.  We have also included machine parameters, such as electron gun temperature, position of its mechanical frequency tuners, current values for the magnets and water flows.

% Discuss results
The red line in Fig.~\ref{fig:Case2} shows the result of the regression. The model is able to learn the correlations in the data and predict to a high degree of accuracy, decreasing the uncertainty in the long-term data (RMSE) by more than a factor of two in the test set.  This represents a major improvement, one that reduces the uncertainty from the hours-long timescale in Fig.~\ref{fig:Stability} to the minutes-long stability.

%, and a factor XXXX in the short-term fluctuations.

% Example
Notably, the model was able to predict accurately the outcome of a sudden large phase shift in the TCAV in the test dataset (visible at around 9000 seconds). Such jumps are therefore due to a sudden variation in the settings of our diagnostic device, and not to an actual change in beam TOA. Starting from the model, real temporal shifts can be isolated from the apparent. Indeed, once the system correlations have been learned from the training dataset, all the coefficients that would contribute to a beam movement on the screen but not necessarily to a change in TOA can be removed. 

% Discuss residual stability and metrics, drift
These improvements, while already impressive, neglect some of the nuances of a drifting, time-series dataset. This is related to the difference between the RMSE quantity, which simply represents the rms of the error of the linear regression prediction from the actual measurement and the standard deviation, which measures the difference from a constant reference value. The choice of reference, in fact, is critical when comparing the performance of the virtual diagnostic to the nominal use-case.  In a nominal case, with traditional feedback, the approach is to reduce the drifts in the system until they can be neglected.  The uncertainty is therefore just the statistical dispersion relative to the given reference point. In other words, if destructive measurements are turned off, the beam is assumed to have the same properties as the last-measured beam.  In a drifting system, on the other hand, if one uses the last measured point as a reference and computes the RMS with respect to that reference, as the time progresses, the uncertainty in the TOA also increases. Using the linear regression virtual diagnostic offers a way to compensate these drifts; there is little to no degradation in the uncertainty as a function of time since the last measurement. 

% Case 2 Sensitivity Analysis
\begin{figure}[h!]
    \begin{center}
  \includegraphics[scale=0.5]{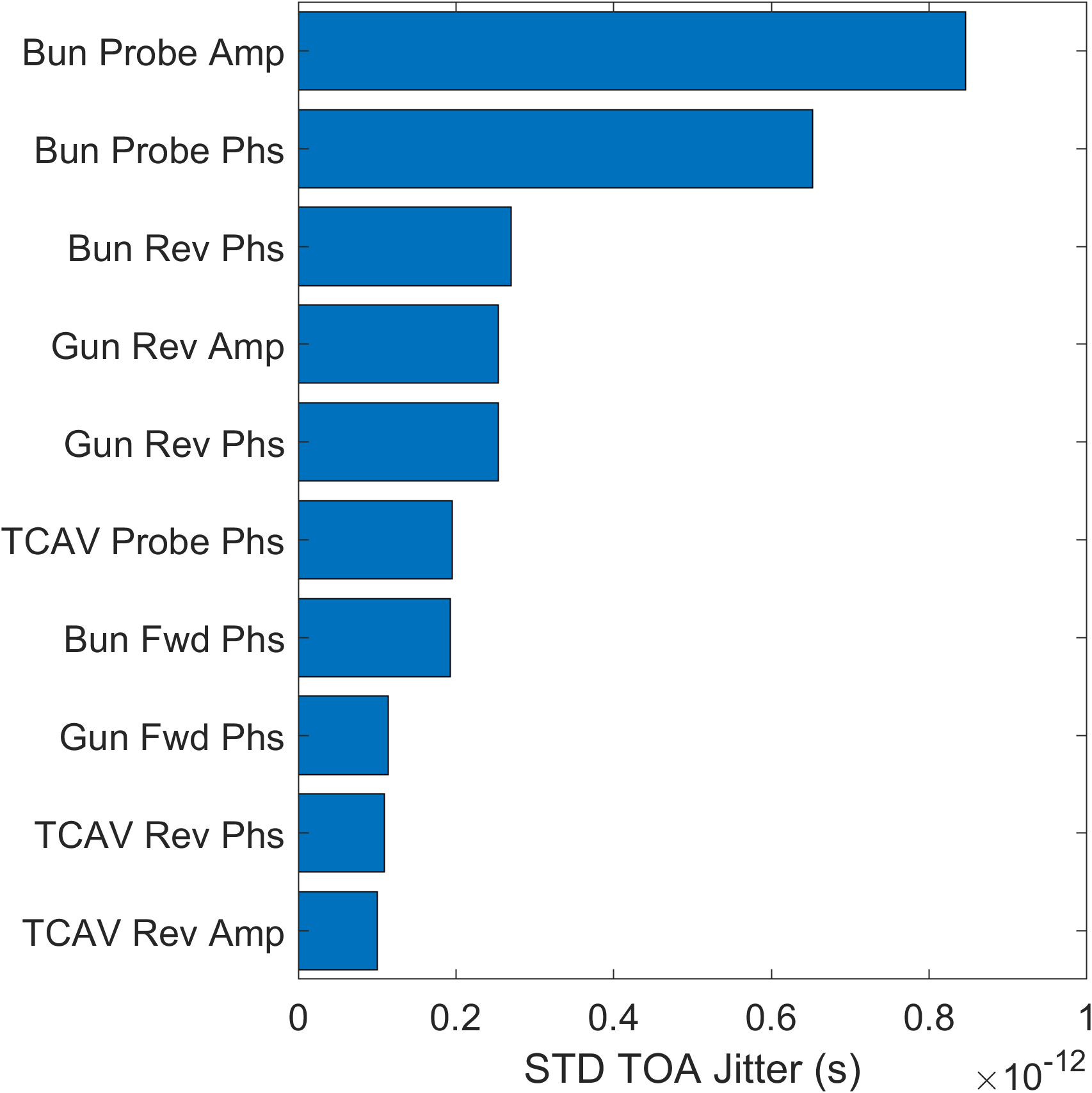}
  \end{center}
  \caption{Top 10 model predictors and the associated TOA movement with 1 STD movement in the validation set.}
  \label{fig:Sensitivity}
\end{figure}

% Discuss sensitivity analysis
As described in the Section \ref{sec:linearmodel}, the impact of each predictor on the overall time of arrival estimate can be extracted from the model. In Fig. \ref{fig:Sensitivity} the top predictors' impacts (one standard deviation of variation converted to TOA prediction using the corresponding $\beta$ from Eq. \ref{eq:LR1}) are reported.  This can be helpful for several reasons, including 1) to see if conventional feedback systems can be better tuned and 2) to see if the perceived TOA variation is due to measurement uncertainty (i.e. the TCAV measuring the TOA is jittering) or if the TOA is actually moving.  Although the effect of the TCAV is significant, the dominating contribution is that of the buncher (Bun in Fig. \ref{fig:Sensitivity}), meaning that the TOA is actually moving, despite conventional feedback systems.  It also suggests that these conventional feedback systems could be improved for the buncher cavity.

% Frame results
These results show how the combination of a linear-regression-based model and time-aligned data can help enhancing the performance of traditional feedback systems. 

\subsection{Energy Stamping}
\label{sec:energystamping}
% Pivot to energy stamping
A similar approach can be used to obtain very accurate predictions of the electron beam energy. To showcase this capability, we make use of separate beamline settings. In particular, the electron beam is transported into the UED line and measured at the VS2 screen (Fig. \ref{fig:HiRES}) after acceleration by the electron gun, while the RF bunching cavity (RF2) is left off for simplicity of interpretation.  We acquired 3 hours of data for the two different cases of stabilized and unstabilized accelerating field in the gun (using the active LLRF PID stabilization loop mentioned earlier).

\begin{figure}[h!]
    \begin{center}
  \includegraphics[scale=0.6]{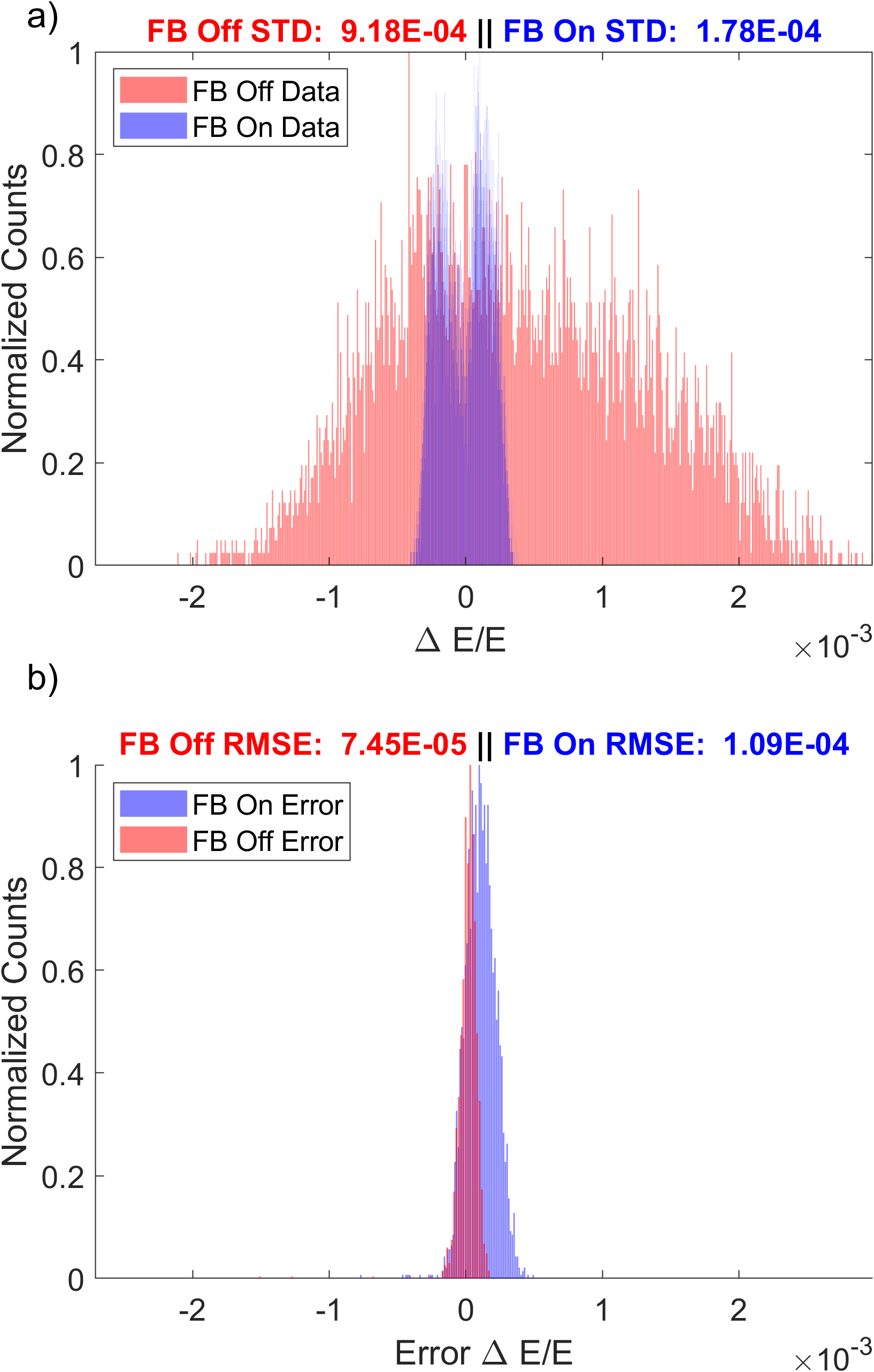}
  \end{center}
  \caption{Linear regression predictions with and without traditional PID-type feedback (FB) engaged.  Top: The variation from the mean of the training data is shown, after conversion to relative energy deviation.  Bottom: the validation errors are shown.  Note that the FB off case shows greater improvement than the FB on case on both an absolute and a relative scale.}
  \label{fig:Case1}
\end{figure}

% Show results
Results of energy stability measurements in the two cases are shown in the histogram of Fig.~\ref{fig:Case1}a. The effect of the fast feedback in stabilizing the energy is evident, with RMS relative energy stability going from approximately $10^{-3}$ to $2\times10^{-4}$ over a 3-hour run. Nevertheless, a clear structure is evident in the stabilized case, with a double peaked distribution of unknown cause, 
suggesting even better performance may be achieved.

The application of our linear regression model to both scenarios results in the histograms of Fig.~\ref{fig:Case1}b. Here we plot the residual error left after comparing the model predictions with the measured ground truth. We can make a few of observations: first, the application of the model increases the precision with which we can assert the energy of each electron beam, by a tenfold factor for feedback-off, and by a small factor in the feedback-on case. 
Second, the residual error distribution is now much closer to a Gaussian, to be expected when only random noise is left, and nothing else can be learned from the system. This is therefore a first indication that our model is close to optimal. Third, the final RMSE  is practically identical in the stabilized and unstabilized case. This last point is quite interesting, since it may suggest that in the not-so-far future with increased computing power and increased network speeds and bandwidths, hardware and software may contribute equally to the control and stabilization of particle beams. 

% Case 1 results

% Comparison of cases

Using a matrix formalism for the linear transport in longitudinal phase space, one can compare the virtual diagnostics results for the energy and time of arrival measurements.  At the TCAV screen,  
\begin{equation}
\Delta t=\left( \frac{R_{\mathrm{56,gb}}}{h \cdot R_{\mathrm{56,gb}}+1}+R_{\mathrm{56,bs}} \right) \frac{\Delta E}{E}
\label{eq:R56_buncher}
\end{equation}
where $R_{\mathrm{56,gb}}$ and $R_{\mathrm{56,bs}}$ are related to the  drift distances from the gun to the buncher and from the buncher to screen.  $h$ is the $R_{65}$ term associated with a thin lens description of the buncher cavity. For HiRES, this can be expressed in engineering units to be approximately:
\begin{equation}
\Delta t \textrm{ [ns]} = 4.3415\frac{\Delta E}{E}
\end{equation}
where $\Delta t$ is given in nanoseconds. Using an uncertainty from the energy stamping virtual diagnostic of about $7.45 \times 10^{-5}$ (see the feedback off case of Fig. \ref{fig:Case1}), we find a value comparable to that of the time stamping virtual diagnostic, when considering the limitations discussed below. 

%\section{Discussion}
\subsection{Limitations of the method}
\label{sec:limits}

\begin{figure}[h!]
    \begin{center}
  \includegraphics[scale=0.55]{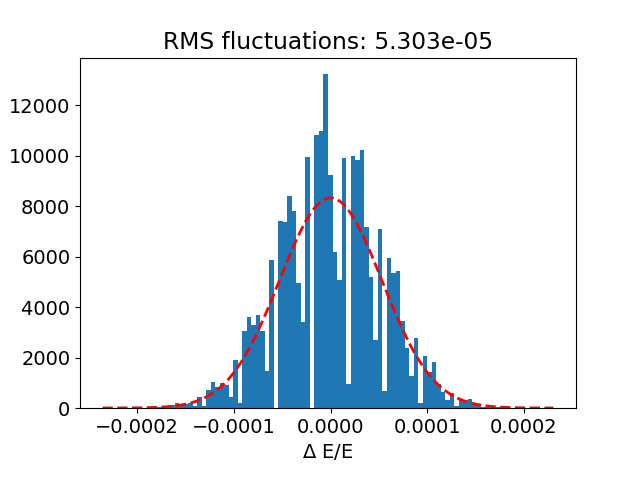}
  \end{center}
  \caption{Measured current fluctuations extrapolated to perceived relative energy fluctuations for a 750 keV electron beam, as in experiment.}
  \label{fig:Dip_Stab}
\end{figure}

There are many possible limitations that leave room to increase the precision of this method beyond what is achieved in this work.  To start, the model used up to now is linear, so any presence of small non-linearity in the system will not be captured. Also, the model may change with time, and therefore some sort of adaptive tuning could be exploited (see~\cite{scheinker_adaptive_2021,scheinker2021adaptive}). Maybe the most important of all limitations, is the accuracy in the measurement of key parameters, and the level of noise in the measured ground truth used for training the model. In this work, we are aiming at final predictions with accuracy at the $10^{-5}$ level, which requires more than 100 dB SNR in measurements of radiofrequency signals. Potentially more dangerous is the requirement on the currents energizing the different magnets in the beamline. Measuring $10^{-5}$ variations on this currents require specialized hardware that it is usually not available for each magnet of the accelerator. Therefore, we perform an experimental sensitivity study and verify the dipole D1 as the one with the highest impact on the beam position on the screen. The current fluctuations driving the magnetic dipole were then measured with high precision by a specialized setup.  
A Danisense DS50ID ultra-stable flux-gate current transducer with a 16-bit digitizer was set up to measure the current provided to the dipole from the CAEN A3620 power supply, when set to a nominal value corresponding to a 750 keV electron beam.  A 66-hour-long measurement of the current was taken, and showed fluctuations on the high $10^{-5}$ level, corresponding to apparent relative energy functions on the of $5\times 10^{-5}$ level.  See Fig. \ref{fig:Dip_Stab} for more details.  
The RMS fluctuations found during this test, although not contextual with beam measurements, are of the same scale of  the residual error we obtained in both cases of Fig.~\ref{fig:Case1}, providing a strong indication on how to improve the precision of the virtual tool. 

% Show dipole fluctuations

\section{Advanced Predictions}
% Segue 
While the above methods are effective, the regression approach presented above relies on the assumption that TOA or energy can be extracted from linear correlations with the predictors.  In this section, an approach is shown that leverages advancements in forecasting to use a Temporal Fusion Tranformer (TFT) architecture to reduce temporal correlation in the residuals and further improve the prediction performance of the model.  

\subsection{Auto-Regressive Models and TFTs}

% %\paragraph{Auto-regressive Models} %\paragraph{LSTM/Sequence Models/Transformers}
Auto-regressive models are a class of models used to represent sequential data that include recurrent neural networks (RNNs), long-term short-term memory networks (LSTMs), and transformers. RNNs estimate the probability $p(y_{t+ 1}| h_t, y_t)$ with a prior consisting either of weights generated from an initialization strategy or previously computed hidden state $h_t$ \cite{rumelhart1985learning}. In practice, representing non-monotonic and complex relationships in sequences requires improving these models by stacking multiple neurons or using bidirectional LSTMs \cite{schuster1997bidirectional,hochreiter1997long,https://doi.org/10.48550/arxiv.1901.03429,sutskever2014sequence}. While still widely used, there are also problems with LSTMs, namely catastrophic forgetting, where successive updates of the memory cell with new data cause the network weights to forget historical data \cite{sodhani2020toward, mccloskey1989catastrophic}. Catastrophic forgetting affects the network's ability to be pre-trained on large datasets prior to fine-tuning on a specific task \cite{pouget2014overcoming}.

%Transformers
Transformers are a member of the class of auto-regressive models that builds upon models such as those described above. Transformers use sub-modules consisting of stacked LSTM neurons as well as novel components such as self-attention in order to train on many datasets or large sequences. While the transformer architecture was originally used in the field of natural language processing, it has subsequently been expanded to a variety of other modalities including time-series forecasting and visual processing \cite{Khan_2022, wolf-etal-2020-transformers}.  Transformers consist of an encoder that maps feature vector $\{x_{1,t},x_{2,t}....x_{n,t}\}$ to a continuous representation $\{z_{1,t},z_{2,t}....z_{n,t}\}$, which is then used by a decoder to generate a sequence of $m$ predictions $\{y_{1,t},y_{2,t}....y_{m,t}\}$. 

%\ Introduce TFTs

TFTs, described in \cite{lim2021temporal}, are a transformer architecture and the current state of the art  for multi-horizon forecasting. At each time step $t$, a context window of length $k$ consisting of past predictors, 
along with the past values of the ground truth, 
are sent to the encoder.  Known machine parameters at the prediction time steps (or horizons) 
are also encoded. 
TFTs contain multiple variable selection networks that reduce network complexity through the modulation of the probability that a given predictor's signal propagates to deeper layers of the network.  This reduces the need for data pre-processing, the impact of noisy variables and prevents over-fitting. 
The TFT decoder includes a multi-headed attention module for long-term temporal pattern recognition.  Through multi-headed attention, weights are learned that reflect the degree to which encoded variables attend or correlate with one another \cite{vaswani2017attention}. These weights are then passed from the decoder to a dense layer that produces the quantile predictions.  
% Advantages of TFT

In summary, the TFT offers an explainable model that predicts a distribution of results instead of a point prediction.  TFTs extract variable importance using both attention and automated variable selection \cite{lim2021temporal}.  Quantile predictions allow confidence of a prediction to be assessed, which is useful for human evaluation and downstream control tasks.

 \subsection{Method}

% Degradation of Error
In offline forecasting tasks, access to past ground truth data allows for the context window to be populated with observations of the target as shown in Eq. \ref{eq:offline_forecast}. A prediction at time step t is given by:
\begin{equation}
    \hat{Y}_t=f(\{\mathbf{X}_{t-k},...,\mathbf{X}_t\},\{Y_{t-k},...,Y_{t-1}\})
    \label{eq:offline_forecast}
\end{equation}
where $\mathbf{X}$ at each time step is a vector of predictors, such as machine parameters.  $Y$ at any given time step is the ground truth data, in this case the beam TOA, where $\hat{Y}$ is the prediction of this quantity.  $k$ is the context window, or history, that the model is given.

In the context of the HIRES virtual diagnostic, the lack of ground truth data during deployment about the beam TOA has to be negotiated once destructive measurements cease.  In such an online or a multi-horizon task, one approach would be to introduce previous predictions recursively as shown in Eq. \ref{eq:multihoizon_forecast}.  
\begin{equation}
    \hat{Y}_t=f(\{\mathbf{X}_{t-k},...,\mathbf{X}_t\},\{\hat{Y}_{t-k},...,\hat{Y}_{t-1}\})
    \label{eq:multihoizon_forecast}
\end{equation}

In this approach, residual bias introduced in the model's estimates would accumulate with each recursive prediction and over thousands of timesteps would become significant. Even without a significant increase in error, error would time-correlated rather than being normally distributed. In order to avoid this problem, once destructive measurements cease, previous ground truth measurements must be replaced with time-independent predictions as shown in Eq. \ref{eq:online_forecast}.  

\begin{equation}
    \hat{Y}_t=f(\{\mathbf{X}_{t-k},...,\mathbf{X}_t\},
    \{g(\mathbf{X}_{t-k}),...,g(\mathbf{X}_{t-1})\})
    \label{eq:online_forecast}
\end{equation}
where $g(x)$ is any time-independent model.  In the work presented herein, the linear regression model as shown in Sec. \ref{sec:timestamping}, as described in Eq. \ref{eq:LR1} is used to replace ground truth data, after training with access to the ground-truth data, as shown in Eq. \ref{eq:offline_forecast}.  Thus, during training, the TFT has access to a context window of long-term trend information in $\mathbf{X}$ and $\mathbf{Y}$ in order to learn from a more complete view of the system's dynamics. Following training, during online application, we utilize accurate estimates of $\mathbf{Y}$ as provided by the linear regression-based estimates $g(\mathbf{X}_t)$ and therefore the approach does not suffer from catastrophic degradation of the predictions after destructive measurements are no longer available.

%\paragraph{Data Dependencies}
The results in this section make use of a TFT implemented in Pytorch Lightning \cite{falcon2019pytorch} with the PyTorch forecasting package \cite{paszke2019pytorch}.  A 75/25 training split identical to that of the linear regression model in Section \ref{sec:onlinePred} was employed with one caveat: for both the training and validation sets, the first \textbf{k} instances (with $k$ being the length of the context window) do not have corresponding predictions. The predictors were the same as those used for the linear regression model in Section \ref{sec:timestamping}.  Finally, a robust normalizer was applied to the data, which scales and centers it with regard to but without transforming the target.

\subsection{Results and Discussion}

 % Advanced Prediction Results
\begin{figure}[h!]
    \begin{center}
  \includegraphics[scale=0.5]{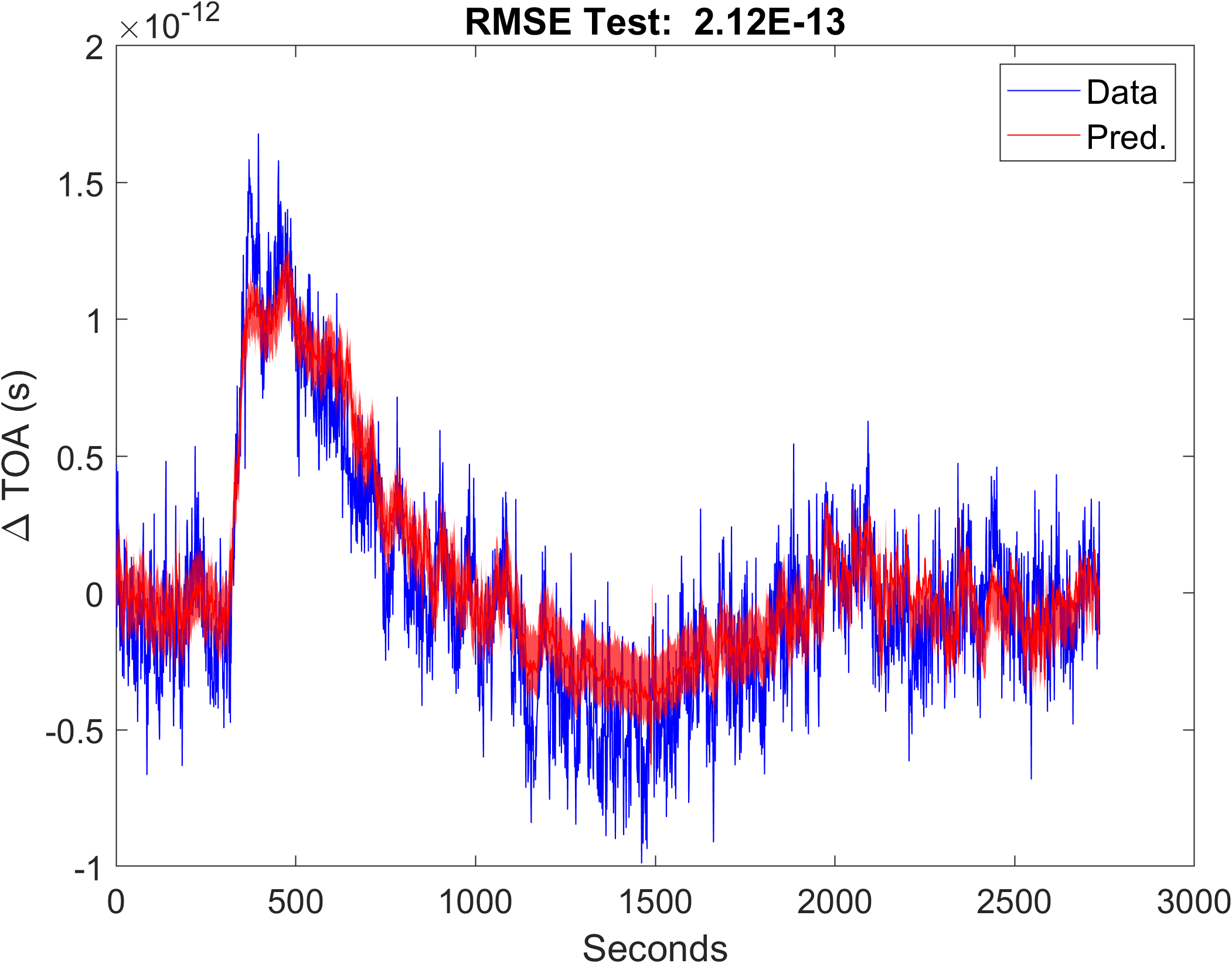}
  \end{center}
  \caption{TFT TOA median and interquartile range (shaded) predictions in the validation set.}
  \label{fig:AdvanceDPrediction}
\end{figure}
\begin{figure}[h]
\includegraphics[scale=0.18]{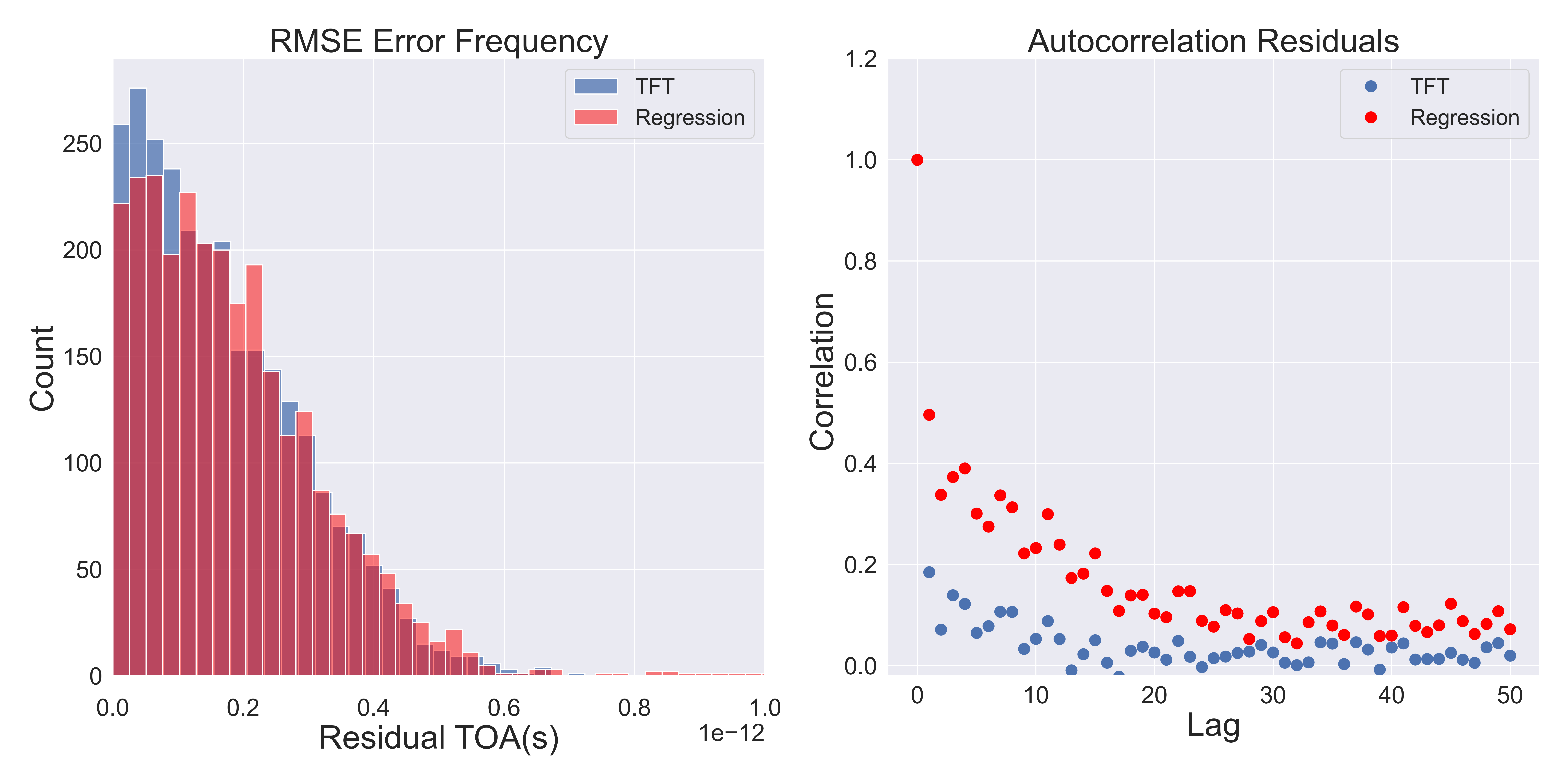}
\caption{ Left: Error histogram comparing the absolute value of the residuals from the linear regression and TFT models.  Note that the error is bunched closer to zero for the TFT.  Right: Residual correlation for linear regression and TFT models.  Note that the use of a TFT predictor reduces the correlation in the residuals between time steps.}
\label{fig:err}
\end{figure}

%\paragraph{Results}
The results shown in Fig. \ref{fig:AdvanceDPrediction} are the quantile predictions trained on ground-truth observations of the beam TOA.  Note that the RMSE is approximately 6\% better than that of linear regression, as the residuals of the model are in general closer to zero, as shown in Fig. \ref{fig:err}.  

This improvement can be explained by noting that despite state-of-the-art stability at HiRES, Fig. \ref{fig:err} demonstrates that the processes causing long-term TOA drift introduce correlations of error over time that could be reduced via the use of a forecasting model.  As shown in Fig \ref{fig:err}, residuals of the TFT show little auto-correlation between time steps relative to the linear regression results. This demonstrates that the approach of incorporating historical machine parameter information despite the challenges of online forecasting detailed above is a reasonable one.  

%\paragraph{Next Steps}
In the future, pre-training on data from different accelerator runs prior to fine-tuning on related data should be investigated to further reduce residual error and allow for increased observation of anomalous states in the machine parameters and their correlated effect on the target beam parameters. Every state-of-the-art use of the Transformer model used in Natural Language Processing since 2015 has relied on pre-training to increase performance \cite{https://doi.org/10.48550/arxiv.1511.01432} and the use of transformers for forecasting would likely benefit from similar methods. The under-fitting (inability for the predictions to capture the full variation of the ground truth beam TOA) exhibited in the plots of Fig. \ref{fig:err} is due to the fact that while relationships between machine parameters and beam TOA have been learned in training, anomalous transitions in machine parameters have not been observed before forcing the model to extrapolate. While anomalous observations are, by definition, rare, a larger number of observations of similar transitions in other experiments would allow for greater predictive power during these periods as well as training with more raw data and possibly with lagged data to capture temporal relationships between changes in the variables. 

Another way to improve accuracy with training data would be training with a greater diversity of examples prior to any aggregation. In addition to learning the between time-step error, learning the variance of the sensor readings within a time-step and training with data taken with the RF cavities' PID controllers disabled would allow the model to learn a more robust embedding space. 

\section{Conclusion}

% Summary of cases
In this work, a novel application of virtual diagnostics has been explored, toward enhancing UED temporal resolution by predicting electron beam TOA – or the main contributor to TOA in this energy regime, beam energy.  Linear-regression-based models can be used to greatly reduce uncertainty in machine parmaeters. For energy stamping, linear-regression-based virtual diagnostics were shown to mitigate the long-term drift to a level comparable to what can be done with the PID feedback loops. For time stamping, linear-regression-based virtual diagnostics were shown to work in concert with traditional feedback to mitigate long-term drift and lower the uncertainty to the short term value of 225 fs.  Further, state-of-the-art forecasting models were applied to mitigate the temporal correlation of residuals of the model predictions, resulting in a nominal reduction in prediction uncertainty.  

% Ways to realize benefits
There are several ways to realize benefits from reducing the uncertainty in prediction error. For example, one could use a virtual diagnostic for feedback, in order to remove the long-term drift. Another method is to make use of the knowledge provided by the virtual diagnostic, without direct feedback. Working under the paradigm of “measurement is easier than control” has been shown to be effective (e.g. \cite{kabra2020mapping,li2018electron}) and has several advantages; rather than working to control further the natural parameter drift of the machine in an already state-of-the-art stability environment, the remaining drift and jitter can be harnessed to improve measurements.  For example, in UED experiments, if for each shot, the virtual diagnostics showcased in this work are applied to retrieve the relative time of arrival within the shot-to-shot error, one would be able to reorder the data using the shot-tag information with a corresponding improvement in the temporal resolution as well as significant reduction of acquisition times.

\begin{acknowledgments}
This material is based partly upon work supported by the U.S. Department of Energy, Office of Science, Office of Workforce Development for Teachers and Scientists, Office of Science Graduate Student Research (SCGSR) program. The SCGSR program is administered by the Oak Ridge Institute for Science and Education for the DOE under contract number DE‐SC0014664.  This work was also partially supported by the DOE Office of Basic Energy Sciences under Contract No. DE-AC02-05CH11231, by the DOE Office of Science, Office of High Energy Physics under contract number 89233218CNA000001 and DE-AC02-05CH11231.  F.C. also acknowledges support from NSF PHY-1549132, Center for Bright Beams.

\end{acknowledgments}

% \bibliography% Produces the bibliography via BibTeX.
% %Control: production of eprint (0) enabled
\providecommand{\noopsort}[1]{}\providecommand{\singleletter}[1]{#1}%

\end{document}